\documentclass[10pt,twocolumn]{article} 

\usepackage{booktabs} 

\usepackage{multirow}
\usepackage{multicol}
\usepackage{balance}
\usepackage{graphicx}
\usepackage{comment}
\usepackage{url}

\usepackage{hyperref}
\usepackage{breakurl}
\usepackage{balance}  
\usepackage{booktabs} 
\usepackage{tabularx}
\usepackage{algorithm, algorithmicx, algpseudocode}
\usepackage{pifont}
\usepackage{xspace}
\usepackage{caption}
\usepackage{subcaption}
\newcommand{\xmark}{\ding{55}}%
\newcommand{\algoname}{CoT\xspace}

\begin{document}

\title{\algoname: Decentralized Elastic Caches for Cloud Environments}

\author{
Victor Zakhary ~~~~ Lawrence Lim ~~~~ Divyakant Agrawal ~~~~ Amr El Abbadi\\
       Department of Computer Science\\
       UC Santa Barbara\\
       Santa Barbara, California, 93106\\
       victorzakhary,lawrenceklim,divyagrawal,elabbadi@ucsb.edu
}

\date{}
\maketitle

\begin{abstract}
Distributed caches are widely deployed to serve social
networks and web applications at billion-user scales.
However, typical workload skew results in load-imbalance
among caching servers. This load-imbalance decreases the 
request throughput and increases the request latency 
reducing the benefit of caching. Recent work
has \textit{theoretically} shown that a 
\textit{small perfect cache} at the front-end has a big 
positive effect on distributed caches load-balance. 
However, determining the cache size and the replacement
policy that achieve near perfect caching at front-end servers is challenging 
especially for dynamically changing and evolving workloads. 
This paper presents \textit{Cache-on-Track }(\algoname), a decentralized, elastic, and 
predictive caching framework for {\bf cloud environments}. \algoname is the answer to
the following question: \textit{What is the necessary front-end cache size
that achieves load-balancing at the caching server side?} 
\algoname proposes a new cache replacement policy specifically tailored for
small front-end caches that serve \textbf{skewed workloads}.
Front-end servers use a heavy hitter tracking
algorithm to continuously track the top-k hot keys. \algoname dynamically 
caches the hottest C keys out of the tracked keys. In addition, each 
front-end server independently monitors its effect on caching servers 
load-imbalance and adjusts its tracker and cache sizes accordingly. 
Our experiments show that \algoname's replacement policy consistently
outperforms the hit-rates of LRU, LFU, and ARC for the same cache size on 
different skewed workloads. Also, \algoname slightly outperforms the 
hit-rate of LRU-2 when both policies are configured with the same tracking 
(history) size. \algoname achieves server size load-balance with 50\% to 
93.75\% less front-end cache in comparison to other replacement policies. 
Finally, our experiments show that \algoname's resizing algorithm 
successfully \textbf{auto-configures} the tracker and cache sizes to achieve back-end load-balance in the presence of workload distribution changes.

\end{abstract}


\section{Introduction} \label{sec:introduction}


Social networks, the web, and mobile applications have attracted 
hundreds of millions of users~\cite{fbInfo,twitterInfo}. These 
users share their relationships and exchange images and videos 
in timely personalized experiences~\cite{bronson2013tao}. 
To enable this real-time experience, the underlying storage systems 
have to provide efficient, scalable, and highly available access to big data. 
Social network users consume several orders of magnitude more data than 
they produce~\cite{atikoglu2012workload}. In addition, a single 
page load requires hundreds of object lookups that need to be 
served in a fraction of a second~\cite{bronson2013tao}. 
Therefore, traditional disk-based storage 
systems are not suitable to handle requests at this scale due to
the high access latency of disks and I/O throughput 
bounds~\cite{zhang2015memory}.

To overcome these limitations, distributed caching services have been 
widely deployed on top of persistent storage in order to 
efficiently serve user requests at scale~\cite{zakhary2017caching}.
Distributed caching systems such as Memcached~\cite{memcached} and 
Redis~\cite{redis} are widely adopted by cloud service providers such 
as Amazon ElastiCache~\cite{elasticache} and Azure Redis Cache~\cite{azureRedis}. 
These caching services offer significant latency and throughput improvements to 
systems that directly access the persistent storage layer. Redis and Memcached use consistent 
hashing~\cite{karger1997consistent} to distribute keys among several caching 
servers.
Although consistent hashing ensures a fair distribution of the number 
of keys assigned to each caching shard, it does not consider the workload 
per key in the assignment process. Real-world workloads are typically skewed 
with few keys being significantly hotter than other 
keys~\cite{huang2014characterizing}. This skew causes load-imbalance among 
caching servers.


Load imbalance in the caching layer can have significant impact on
the overall application performance. In particular, it may cause drastic increases in
the latency of operations at the tail end of the access frequency 
distribution~\cite{hong2013understanding}.
In addition, the average throughput decreases and the average
latency increases when the workload skew increases~\cite{cheng2015memory}. This increase 
in the average and tail latency is amplified for real workloads when operations are 
executed in chains of dependent data objects~\cite{lu2016snow}. A single Facebook 
page-load results in retrieving hundreds of objects in multiple rounds of
data fetching operations~\cite{nishtala2013scaling, bronson2013tao}. Finally,
solutions that equally overprovision the caching layer resources to handle the 
most loaded caching server suffer from \textbf{resource under-utilization} in
the least loaded caching servers.

Various approaches have been proposed to solve the 
load-imbalance problem using centralized load 
monitoring~\cite{adya2016slicer, wu2019autoscaling}, server side load 
monitoring~\cite{hong2013understanding}, or front-end load 
monitoring~\cite{fan2011small}. Adya et al.~\cite{adya2016slicer}
propose Slicer that separates the data serving plane from the control plane. 
The control plane is a centralized system component that collects metadata 
about shard accesses and server workload. It periodically runs an 
optimization algorithm that decides to redistribute, repartition, or 
replicate slices of the key space to achieve better back-end load-balance.
Hong et al.~\cite{hong2013understanding} use a distributed server
side load monitoring to solve the load-imbalance problem. 
Each back-end server independently tracks its hot keys and decides to 
distribute the workload of its hot keys among other back-end servers.
Solutions in~\cite{adya2016slicer,wu2019autoscaling} and~\cite{hong2013understanding}
require the back-end to change the key-to-caching-server mapping
and announce the new mapping to all the front-end servers.
Fan et al.~\cite{fan2011small} use a distributed front-end load-monitoring 
approach. This approach shows that adding a small cache in the front-end 
servers has significant impact on solving the back-end load-imbalance. 
Caching the heavy hitters at front-end servers reduces the skew among the 
keys served from the caching servers and hence achieves better back-end 
load-balance. Fan et al. {\em theoretically} show 
through analysis and simulation that a small \textit{perfect cache}
at each front-end solves the back-end load-imbalance problem. However,
perfect caching is practically hard to achieve. 
Determining the cache size and the replacement policy that
achieve near perfect caching at the front-end for dynamically changing and 
evolving workloads is challenging. 


In this paper, we propose \textbf{Cache-on-Track }(\algoname); a 
\textbf{decentralized}, \textbf{elastic}, and \textbf{predictive} heavy 
hitter caching at front-end servers. \algoname proposes a new 
cache replacement policy specifically tailored for small front-end caches 
that serve \textbf{skewed workloads}.
\algoname 
uses a small front-end cache to solve back-end load-imbalance
as introduced in~\cite{fan2011small}. However, \algoname does not assume 
perfect caching at the front-end.
\algoname uses the space saving 
algorithm~\cite{metwally2005efficient} to track the \textbf{top-k} 
\textit{heavy hitters}. The tracking information allows \algoname to {\em 
cache} the exact top C hot-most keys out of the approximate top-k tracked 
keys preventing cold and noisy keys from the long tail to replace hot keys 
in the cache. \algoname is decentralized in the sense that each front-end 
independently determines its hot key set based on the key access 
distribution served at this specific front-end. This allows \algoname
to address back-end load-imbalance without introducing single points
of failure or bottlenecks that typically come with centralized solutions. In
addition, this allows \algoname to scale to thousands of front-end servers, 
a common requirement of social network and modern web applications.
\algoname is elastic in that each front-end uses its 
local load information to monitor its contribution to the 
back-end load-imbalance. Each front-end elastically adjusts 
its tracker and cache sizes to reduce the load-imbalance caused by 
this front-end. In the presence of workload changes, \algoname dynamically
adjusts front-end tracker to cache ratio in addition to both the 
tracker and cache sizes to eliminate any back-end load-imbalance.

In traditional architectures, memory sizes are static and 
caching algorithms strive to achieve the best usage of all the 
available resources. However, in a cloud setting where there are 
theoretically infinite memory and processing resources and cloud instance 
migration is the norm,  cloud end-users aim to achieve their SLOs while
reducing the required cloud resources and thus 
decreasing their monetary deployment costs. \algoname's main goal is
to reduce the necessary front-end cache size at
each front-end to eliminate server-side load-imbalance. Reducing front-end 
cache size is crucial for the following reasons: 1) it reduces
the monetary cost of deploying front-end caches. For this, we quote David
Lomet in his recent works~\cite{lomet2019data,lomet2018cost,lomet2018caching} where he shows that
cost/performance is usually more important than sheer performance:
\textit{"the argument here is not that there is insufficient main memory to 
hold the data, but that there is a less costly way to manage data."}. 2) In 
the presence of data updates and when data consistency is a requirement, 
increasing front-end cache sizes significantly increases the cost of the 
data consistency management technique. Note that social networks and modern
web applications run on thousands of front-end servers. Increasing
front-end cache size not only multiplies the cost of deploying bigger
cache by the number of front-end servers, but also increases several
costs in the consistency management pipeline including a) the cost
of tracking key incarnations in different front-end servers and b)
the network and processing costs to propagate updates to front-end
servers. 3) Since the workload is skewed, our experiments clearly demonstrate that the relative benefit of adding more front-end
cache-lines, measured by the average cache-hits per 
cache-line and back-end load-imbalance reduction, drastically decreases as front-end cache sizes increase.

\algoname's resizing algorithm dynamically increases or decreases 
front-end allocated memory in response to dynamic workload changes. 
\algoname's dynamic resizing algorithm is valuable in different cloud
settings where
1) all front-end servers are deployed in the same datacenter and obtain the 
same dynamically evolving workload distribution, 2) all front-end servers are
deployed in the same datacenter but obtain different dynamically evolving
workload distributions, and finally 3) front-end servers are deployed at 
different \textit{edge-datacenters} and obtain different dynamically evolving
workload distributions. In particular, \algoname aims to capture local 
trends from each individual front-end server perspective. In social network 
applications, front-end servers that serve different geographical regions 
might experience different key access distributions and different local 
trends (e.g., \#miami vs. \#ny). Similarly, in large scale data processing 
pipelines, several applications are deployed on top of a shared caching 
layer. Each application might be interested in different partitions of the 
data and hence experience different key access distributions and local 
trends. While \algoname operates on a \textit{fine-grain key} level at 
front-end servers, solutions like Slicer~\cite{adya2016slicer} operate on 
coarser grain slices or shards at the caching servers. Server side solutions
are complementary to \algoname. Although capturing local trends alleviates 
the load and reduces load-imbalance among caching servers, other factors can result in load-imbalance and 
hence using server-side load-balancing, e.g., Slicer, might still be 
beneficial. 

We summarize our contributions in this paper as follows.
\begin{itemize}
    \item Cache-on-Track (\algoname) is a decentralized, elastic, and predictive front-end caching framework that reduces back-end load-imbalance and
    improves overall performance.
    
    \item \algoname dynamically minimizes the required front-end cache size to achieve
    back-end load-balance. \algoname's built-in elasticity is a key novel advantage over other replacement policies.
    
    \item Extensive experimental studies that compare \algoname's
    replacement policy to both traditional as well as state-of-the-art replacement
    policies, namely, LFU, LRU, ARC, and LRU-2. The experiments demonstrate that \algoname 
    achieves server size load-balance for different workload with \textbf{50\% to 93.75\%} less front-end cache in comparison to other replacement policies. 
    
    \item The experimental study demonstrates that \algoname 
    successfully auto-configures its tracker and cache sizes to achieve back-end 
    load-balance.
    
    \item In our experiments, we found a \textit{bug} in
    YCSB's~\cite{cooper2010benchmarking} ScrambledZipfian workload generator. This 
    generator generates workloads that are significantly less-skewed than the promised 
    Zipfian distribution.
\end{itemize}

The rest of the paper is organized as follows. In 
Section~\ref{sec:model}, the system and data models are explained.
In Section~\ref{sec:client_alt}, we motivate \algoname by presenting the main advantages and
limitations of using LRU, LFU, ARC, and LRU-k caches at the front-end. We present the 
details of \algoname in Section~\ref{sec:cot}. In 
Section~\ref{sec:evaluation}, we evaluate the performance and the overhead of 
\algoname. The related work is discussed in Section~\ref{sec:related_work} and the paper is concluded in Section~\ref{sec:conclusion}.

\section{System and Data Models} \label{sec:model}

\begin{figure}[ht!]
    \centering
    \includegraphics[width=0.5\textwidth]{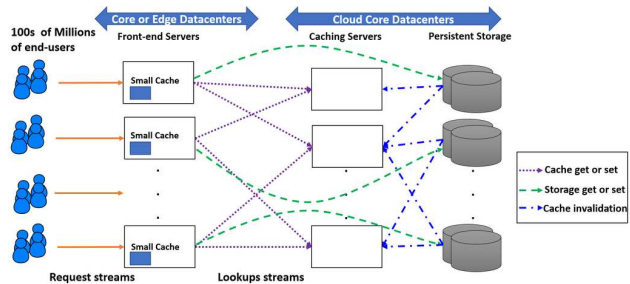}
    \caption{Overview of the system architecture.
    }
        \label{fig:architecture}
\end{figure}

This section introduces the system and data access models. Figure~\ref{fig:architecture}
presents the system architecture where user-data is stored in a distributed back-end
storage layer in the cloud. The back-end storage layer consists of a distributed in-memory 
caching layer deployed on top of a distributed persistent storage layer. The caching layer 
aims to improve the request latency and system throughput and to alleviate the load on the 
persistent storage layer. As shown in Figure~\ref{fig:architecture}, hundreds of millions of end-users send
streams of page-load and page-update requests to thousands of stateless front-end 
servers. These front-end servers are either deployed in the same core datacenter as
the back-end storage layer or distributed among other core and edge datacenters
near end-users. Each end-user request 
results in hundreds of data object lookups and updates served from
the back-end storage layer. According to Facebook Tao~\cite{bronson2013tao}, 99.8\% of 
the accesses are reads and 0.2\% of them are writes. Therefore, the storage system 
has to be \textbf{read optimized} to efficiently handle end-user requests at scale.

The front-end servers can be viewed as the clients of the back-end storage layer. 
We assume a typical key/value store interface between the front-end servers and 
the storage layer. The API consists of the following calls:
\begin{itemize}
\item \textit{v = get(k)} retrieves value \textit{v} corresponding to key \textit{k}.
\item \textit{set(k, v)} assigns value \textit{v} to key \textit{k}.
\item \textit{delete(k)} deletes the entry corresponding key \textit{k}.    
\end{itemize}
 
Front-end servers use \textit{consistent hashing}~\cite{karger1997consistent} to locate 
keys in the caching layer. Consistent hashing solves the \textit{key discovery} 
problem and reduces key churn when a caching server is added to or removed from the 
caching layer. We extend this model 
by adding an additional layer in the cache hierarchy. As shown in 
Figure~\ref{fig:architecture}, each front-end server maintains a small cache of 
its hot keys. This cache is populated according to the accesses that are served by this 
front-end server.

We assume a client driven caching protocol similar to the protocol implemented by
\textbf{Memcached}~\cite{memcached}. A cache client library is deployed in the front-end servers. 
\textit{Get} requests are initially attempted to be served from the local cache.
If the requested key is in the local cache, the \textit{value} is returned and the 
request is marked as served. Otherwise, a \textit{null} value is returned and 
the front-end has to request this key from the caching layer {\bf at the back-end storage layer}. 
If the key is cached in the caching layer, its value is returned to the front-end. Otherwise, 
a \textit{null} value is returned and the front-end has to request this key 
from the persistent storage layer and upon receiving the corresponding value, the front-end inserts the 
value in its front-end local cache and in the server-side caching layer as well. As in~\cite{nishtala2013scaling}, a 
\textit{set}, or an update, request invalidates the key in both the local cache and the 
caching layer. Updates are directly sent to the 
persistent storage, local values are set to null, and delete requests are sent to 
the caching layer to invalidate the updated keys. The Memcached client driven
approach allows the deployment of a \textit{stateless} caching layer. As requests are
driven by the client, a caching server does not need to maintain the state of
any request. This simplifies scaling and tolerating failures at the caching layer.
Although, we adopt the Memcached client driven request handling protocol, our
model works as well with write-through request handling protocols.

Our model is not tied to any replica consistency model. Each key can have multiple 
incarnations in the storage layer and the caching layer. Updates can be synchronously
propagated if \textit{strong consistency} guarantees are needed or 
asynchronously propagated if \textit{weak consistency} guarantees suffice. Achieving strong
consistency guarantees among replicas of the same object has been widely studied 
in~\cite{cheng2015memory,hong2013understanding}.
Ghandeharizadeh et 
al.~\cite{ghandeharizadeh2019rejig, ghandeharizadeh2019design} propose several
complementary techniques to \algoname to deal with consistency
in the presence of updates and configuration changes. These techniques can easily be adopted
in our model according to the application requirements. We understand
that deploying an additional vertical layer of cache increases potential 
data inconsistencies and hence increases update propagation and 
synchronization overheads. Therefore, our goal in this paper is to 
reduce the front-end cache size in order to limit the inconsistencies
and the synchronization overheads that result from deploying front-end 
caches, while maximizing their benefits.

\section{Front-end Cache Alternatives} \label{sec:client_alt}

Fan et al.~\cite{fan2011small} show that a small cache in the front-end servers
has big impact on the caching layer load-balance. Their analysis assumes 
perfect caching in front-end servers for the hottest keys. A \textbf{perfect cache} of $C$ cache-lines 
is defined such that accesses for the $C$ hot-most keys always hit the cache while 
other accesses always miss the cache. However, the perfect caching assumption is 
impractical especially for dynamically changing and evolving workloads. Different 
replacement policies have been developed to approximate perfect caching for different 
workloads. In this section, we discuss the workload assumptions and various client 
caching objectives. This is followed by a discussion 
of the advantages and limitations of common caching replacement policies such as 
Least Recently Used (LRU), Least Frequently Used (LFU), Adaptive Replacement
Cache (ARC~\cite{megiddo2003arc}) and LRU-k~\cite{o1993lruk}.

\textbf{Workload assumptions:} Real-world workloads are typically skewed with few 
keys being significantly hotter than other keys~\cite{huang2014characterizing}. 
Zipfian distribution is a common example of a key hotness distribution.  However, key hotness can follow different distributions 
such as \textit{Gaussian} or different variations of Zipfian~\cite{breslau1999web,
guo2008stretched}. In this paper, we assume skewed workloads with periods of 
stability (where hot keys remain hot during these periods).

\textbf{Client caching objectives:} Front-end servers construct their perspective of 
the key hotness distribution based on the requests they serve. Front-end servers
aim to achieve the following caching objectives:
\begin{itemize}

\item The cache replacement policy should prevent cold keys from 
replacing hotter keys in the cache.

\item Front-end caches should adapt to the changes in the workload. In particular, front-end
servers should have a way to retire hot keys that are no longer accessed. In 
addition, front-end caches should have a mechanism to expand or shrink their local 
caches in response to changes in workload distribution. For example, front-end 
servers that serve uniform access distributions should dynamically shrink their 
cache size to \textit{zero} since caching is of no value in this situation. On the other hand, front-end servers that serve highly 
skewed Zipfian (e.g., s = 1.5) should dynamically expand 
their cache size to capture all the hot keys that cause load-imbalance among the back-end 
caching servers.

\end{itemize}

A popular policy for implementing client caching is the LRU replacement policy.
Least Recently Used (LRU) costs O(1) per access and caches keys based on their recency of access. This 
may allow cold keys that are recently accessed to replace hotter cached keys. 
Also, LRU cannot distinguish well between frequently and infrequently accessed 
keys~\cite{lee2001lrfu}. For example, this access sequence (A, B, C, D, A, B, C, E, 
A, B, C, F, ...) would always have a cache miss for an LRU cache of size 3. 
Alternatively, Least Frequently Used (LFU) can be used as a replacement policy. 
LFU costs $O(log(C))$ per access where $C$ is the cache size. LFU is typically 
implemented using a min-heap and allows cold keys to replace hotter keys at the 
root of the heap. Also, LFU cannot distinguish between old references and recent 
ones. For example, this access sequence (A, A, B, B, C, D, E, C, D, E, C, D, E 
....) would always have a cache miss for an LFU cache of size 3 except for the $2^{nd}$ and 
$4^{th}$ accesses. This means that LFU cannot adapt to changes in 
workload. Both LRU and LFU are limited in their knowledge to the content of the 
cache and cannot develop a wider perspective about the hotness distribution outside 
of their static cache size. Our experiments in Section~\ref{sec:evaluation} show that
replacement policies that track more keys beyond their cache sizes (e.g., ARC, LRU-k, and \algoname)
beat the hit-rates of replacement policies that have no access information of keys beyond their cache
size especially for periodically stable skewed workloads.

Adaptive Replacement
Cache (ARC)~\cite{megiddo2003arc} tries to realize the benefits of both LRU and LFU policies by maintaining two caching lists: one for \emph{recency} and one
for \emph{frequency}. ARC dynamically changes the number of cache-lines allocated for each list to
either favor recency or frequency of access in response to workload changes. In addition, 
ARC uses shadow queues to track more keys beyond the cache size. This helps ARC to maintain 
a broader perspective of the access distribution beyond the cache size. ARC is designed to find 
the fine balance between recent and frequent accesses. As a result, ARC pays the cost of caching 
every new cold key in the recency list evicting a hot key from the frequency list. This cost is
significant especially when the cache size is much smaller than the key space and the workload
is skewed favoring frequency over recency.

LRU-k tracks the last k accesses for each key in the cache, in addition to a pre-configured
fixed size history that include the access information of the recently evicted keys from the
cache. New keys replace the key with the least recently $k^{th}$ access in the cache. The evicted key
is moved to the history, which is typically implemented using a LRU like queue. LRU-k is a suitable 
strategy to mock perfect caching of periodically stable skewed workloads when its cache and history sizes 
are perfectly pre-configured for this specific workload. However, due to the lack of LRU-k's dynamic 
resizing and elasticity of both its cache and history sizes, we choose to introduce \algoname that is
designed with native resizing and elasticity functionality. This functionality allows \algoname to adapt
its cache and tracker sizes in response to workload changes.

%

\section{Cache on Track (\algoname)}\label{sec:cot}

Front-end caches serve two main purposes: \textit{1) decrease the load on the back-end caching 
layer} and \textit{ 2) reduce the load-imbalance among the back-end caching servers}. \algoname focuses
on the latter goal and considers back-end load reduction a complementary side effect. 
\algoname's design philosophy is to track more keys beyond the cache size. This tracking serves as a
filter that prevents cold keys from populating the small cache and therefore, only hot keys can populate
the cache. In addition, the tracker and the cache are dynamically and adaptively resized to ensure that
the load served by the back-end layer follows a load-balance target.

The idea of tracking more keys beyond the cache size has been widely used in replacement
policies such as 2Q~\cite{johnson1994x3}, MQ~\cite{zhou2001multi}, LRU-k~\cite{o1993lruk, o1999lruk2}, 
ARC~\cite{megiddo2003arc}, and in other works like Cliffhanger~\cite{cidon2016cliffhanger} 
to solve other cache problems. Both 2Q and MQ use multiple LRU queues to overcome the weaknesses 
of LRU of allowing cold keys to replace warmer keys in the cache. 
Cliffhanger uses
shadow queues to solve a different problem of memory allocation among cache blobs. 
All these policies
are desgined for fixed memory size environments. However, in a cloud environment
where elastic resources can be requested on-demand, a new cache replacement policy
is needed to take advantage of this elasticity.

\algoname presents a new cache replacement policy that uses a \textit{shadow
heap} to track more keys beyond the cache size. Previous works 
have established the efficiency of heaps in tracking frequent items~\cite{metwally2005efficient}.
In this section, we explain how \algoname uses tracking beyond the cache 
size to achieve the caching objectives listed in Section~\ref{sec:client_alt}. 
In particular, \algoname answers the following 
questions: 1) \textit{how to prevent cold keys from replacing hotter keys in the cache?}, 2) 
\textit{how to reduce the required front-end cache size that achieves lookup load-balance?}, 
3) \textit{how to adaptively resize the cache in response to changes in
the workload distribution?}  and finally 
4) \textit{how to dynamically retire old heavy hitters?}.

First, we develop the notation in Section~\ref{sub:notation}. Then, we explain 
the space saving tracking algorithm~\cite{metwally2005efficient} in 
Section~\ref{sub:space_saving}. \algoname uses the space saving algorithm 
to track the approximate \textit{top-k} keys in the lookup stream. In
Section~\ref{sub:local_cache_management}, we extend the space saving algorithm
to capture the exact top \textit{C} keys out of the approximately tracked
\textit{top-k} keys. \algoname's cache replacement policy dynamically captures and 
caches the exact top C keys thus preventing cold keys from replacing hotter keys 
in the cache. \algoname's adaptive cache resizing algorithm is presented in 
Section~\ref{sub:resizing_algorithm}. \algoname's resizing algorithm
exploits the elasticity and the migration flexibility of the cloud and 
minimizes the required front-end memory size to achieve back-end load-balance.
Section~\ref{sub:expanding} explains how \algoname expands and shrinks front-end tracker and cache 
sizes in response to changes in workload.

\subsection{Notation}\label{sub:notation}

\begin{table}[!ht]
\centering
\begin{tabularx}{\columnwidth}{ | l | X |}
\hline
$S$ & key space \\ \hline
$K$ & number of tracked keys at the front-end \\ \hline
$C$ & number of cached keys at the front-end \\ \hline
$h_k$ & hotness of a key \textit{k} \\ \hline
$k.r_c$ & read count of a key k \\ \hline
$k.u_c$ & update count of a key k \\ \hline
$r_w$ & the weight of a read operation\\ \hline
$u_w$ & the weight of an update operation \\ \hline
$h_{min}$ &  the minimum key hotness in the cache\\ \hline
$S_{k}$ & the set of all tracked keys \\ \hline
$S_{c}$ & the set of tracked and cached keys \\ \hline
$S_{k-c}$ & the set of tracked but not cached keys \\ \hline
$I_c$ &  the current local lookup load-imbalance \\ \hline
$I_t$ &  the target lookup load-imbalance \\ \hline
$\alpha$ & the average hit-rate per cache-line\\ \hline

\end{tabularx}
\caption{Summary of notation.}
\label{table:notation}
\end{table}

The key space, denoted by $S$, is assumed to be large in the scale of trillions
of keys. Each front-end server maintains a cache of size $C <<< S$. The set of 
cached keys is denoted by $S_c$. To capture the 
hot-most $C$ keys, each front-end server tracks $K > C$ keys. The set
of tracked key is denoted by $S_k$. Front-end servers cache the 
hot-most $C$ keys where $S_c \subset S_k$. A key hotness $h_k$ is determined using 
the dual cost model introduced in~\cite{dasgupta2017caching}. In this model, read 
accesses increase a key hotness by a read weight $r_w$ while update accesses 
decrease it by an update weight $u_w$. As update accesses cause
cache invalidations, frequently updated keys should not be cached 
and thus an update access decreases key hotness. For each tracked key, the
read count $k.r_c$ and the update count $k.u_c$ are maintained to capture the number
of read and update accesses of this key. Equation~\ref{equ:hotness} shows how
the hotness of key $k$ is calculated.

\begin{equation}
h_k = k.r_c \times r_w - k.u_c \times u_w
\label{equ:hotness}
\end{equation}

$h_{min}$ refers to the minimum key hotness in the cache. $h_{min}$ splits 
the tracked keys into \textit{two} subsets: 1) the set of tracked and cached keys 
$S_{c}$ of size $C$ and 2) the set of tracked but not cached keys $S_{k-c}$ of 
size $K-C$. The current local load-imbalance among caching servers lookup load
is denoted by $I_c$. $I_c$ is a local variable at each front-end that determines the current contribution of this front-end to the back-end load-imbalance. $I_c$ is defined as the workload ratio between the most loaded back-end server and the least loaded back-end server as observed at a front-end server. For example, if a front-end server sends, during an epoch, a maximum of 5K key lookups to some back-end server and, during the same epoch, a minimum of 1K key lookups to another back-end server then $I_c$, at this front-end, equals $5$. $I_t$ is the 
target load-imbalance among the caching servers. $I_t$ is the only input parameter set by the system 
administrator and is used by front-end servers to dynamically adjust their cache
and tracker sizes. Ideally $I_t$ should be set close to 1. $I_t = 1.1$ means that back-end load-balance is achieved if the most loaded server observe at most 10\%
more key lookups that the least loaded server. Finally, we define another \textbf{local 
auto-adjusted} variable $\alpha$. $\alpha$ is the average hits 
per cache-line and it determines the quality of the cached keys. $\alpha$ helps 
detect changes in workload and adjust the cache size accordingly. Note that \algoname 
automatically infers the value of $\alpha$ based on the observed workload. Hence, 
the system administrator does not need to set the value of $\alpha$.
Table~\ref{table:notation} summarizes the notation.

\subsection{Space-Saving Hotness Tracking Algorithm} \label{sub:space_saving}

We use the \textit{space-saving} algorithm introduced in~\cite{metwally2005efficient} 
to track the key hotness at front-end servers. Space-saving
uses a min-heap to order keys based on their hotness and a hashmap to lookup keys 
in the tracker in \textit{O(1)}. The space-saving algorithm is shown
in Algorithm~\ref{algo:space_saving}. If the accessed key $k$ is not in the 
tracker (Line~\ref{line:1}), it replaces the key with minimum hotness at the root
of the min-heap (Lines~\ref{line:2},~\ref{line:3}, and~\ref{line:4}). The algorithm
gives the newly added key the benefit of doubt and assigns it the hotness of the 
replaced key. As a result, the newly added key gets the opportunity to survive 
immediate replacement in the tracker. Whether the accessed key $k$ was in the tracker 
or is newly added to the tracker, the hotness of the key is updated based on the 
access type according to Equation~\ref{equ:hotness} (Line~\ref{line:update_hotness}) 
and the heap is accordingly adjusted (Line~\ref{line:adjust_heap}).  

\begin{algorithm}[!h]
\caption{The space-saving algorithm: track\_key( key k, access\_type t).}
 \label{algo:space_saving}
\begin{flushleft}
\textbf{State:} $S_k$: keys in the tracker. \newline
\textbf{Input:} (key k, access\_type t)
\end{flushleft}
\begin{algorithmic}[1]
	\If {$k \notin S_k$} \label{line:1}
		\State let $k^{'}$ be the root of the min-heap  \label{line:2}
    	\State replace $k^{'}$ with $k$ \label{line:3}
    	\State $h_k$ := $h_{k^{'}}$  \label{line:4}
  	\EndIf
	\State $h_k$ := update\_hotness(k, t) \label{line:update_hotness}
	\State adjust\_heap(k) \label{line:adjust_heap}
    \State return $h_k$
\end{algorithmic}
\end{algorithm}

\subsection{\algoname: Cache Replacement Policy} \label{sub:local_cache_management}

\algoname's tracker captures the approximate top $K$ hot keys.
Each front-end server should cache the exact top $C$ keys
out of the tracked $K$ keys where 
$C < K$. The exactness of the top $C$ cached keys is considered with respect to the 
approximation of the top $K$ tracked keys.
Caching the exact top $C$ keys prevents cold and noisy keys from replacing hotter
keys in the cache and achieves the first caching objective. To determine the exact top
$C$ keys, \algoname maintains a cache of size C in a min-heap structure.
Cached keys are partially ordered in the min-heap based on their hotness. The root of 
the cache min-heap gives the minimum hotness, $h_{min}$, among the cached keys. $h_{min}$
splits the tracked keys into \textit{two unordered} subsets $S_{c}$ and $S_{k-c}$
such that:
\begin{itemize}
\item $|S_{c}| = C$ and $\forall_{x \in S_{c}} h_{x} \ge h_{min}$
\item $|S_{k-c}| = K - C$ and $\forall_{x \in S_{k-c}} h_{x} < h_{min}$
\end{itemize}

\begin{figure}[ht!]
    \centering
    \includegraphics[width=\columnwidth]{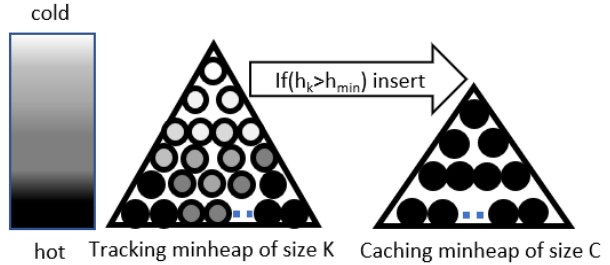}
    \caption{\algoname: a key is inserted to the cache 
    if its hotness exceeds the minimum hotness of the cached
    keys.}
        \label{fig:hotcache}
\end{figure}

\vspace{-10pt}

For every key access, the hotness information of the accessed key 
is updated in the tracker. If the accessed key is cached, its 
hotness information is updated in the cache as well. However, if
the accessed key is not cached, its hotness is compared against 
$h_{min}$. As shown in Figure~\ref{fig:hotcache}, the accessed 
key is inserted into the cache only if its hotness exceeds 
$h_{min}$. Algorithm~\ref{algo:cot_caching} explains the details 
of \algoname's cache replacement algorithm.

\begin{algorithm}[!h]
\caption{\algoname's caching algorithm}
 \label{algo:cot_caching}

\begin{flushleft}
\textbf{State:} $S_k$: keys in the tracker and $S_c$: keys in the cache.\newline
\textbf{Input:} (key k, access\_type t)
\end{flushleft}
\begin{algorithmic}[1]
	\State $h_k$ = track\_key(k, t) as in Algorithm~\ref{algo:space_saving} \label{line:tracker}
	\If {$k \in S_c$} \label{line:cache_check}
		\State let v = access($S_c$, k) // local cache access \label{line:cache_access}
    \Else
        \State let v = server\_access(k) // caching server access \label{line:server_access}
        \If {$h_k > h_{min}$} \label{line:hmin}
            \State insert($S_c$, k, v) // local cache insert \label{line:cache_insert}
        \EndIf
  	\EndIf
	\State return v
\end{algorithmic}
\end{algorithm}

For every key access, the \textit{track\_key} function of 
Algorithm~\ref{algo:space_saving} is called 
(Line~\ref{line:tracker}) to update the tracking information 
and the hotness of the accessed key. Then, a key access is served 
from the local cache only if the key is in the cache 
(Lines~\ref{line:cache_access}). Otherwise, the access is served 
from the caching server (Line~\ref{line:server_access}). Serving 
an access from the local cache implicitly updates the accessed 
key hotness and location in the cache min-heap. If the accessed 
key is not cached, its hotness is compared against $h_{min}$ 
(Line~\ref{line:hmin}). The accessed key is inserted to the local 
cache if its hotness exceeds $h_{min}$ 
(Line~\ref{line:cache_insert}). This happens only if there is a
tracked but not cached key that is hotter than one of the cached keys. 
Keys are inserted to the cache together with their tracked hotness 
information. Inserting keys into the cache follows 
the LFU replacement policy. This implies that a local cache insert 
(Line~\ref{line:cache_insert}) would result in the replacement of 
the coldest key in the cache (the root of the cache heap) if the local cache is full.

\subsection{\algoname: Adaptive Cache Resizing}\label{sub:resizing_algorithm}
This section answers the following questions:  \textit{how to reduce the necessary
front-end cache size that achieves front-end lookup load-balance?} \textit{How to shrink 
the cache size when the workload's skew decreases?} and \textit{How to detect changes in 
the set of hot keys?} As explained in Section~\ref{sec:introduction}, Reducing the
front-end cache size decreases the front-end cache
monetary cost, limits the overheads of data consistency
management techniques, and maximizes the benefit of 
front-end caches measured by the average cache-hits per 
cache-line and back-end load-imbalance reduction.

\begin{figure}[ht!]
    \centering
    \includegraphics[width=\columnwidth]{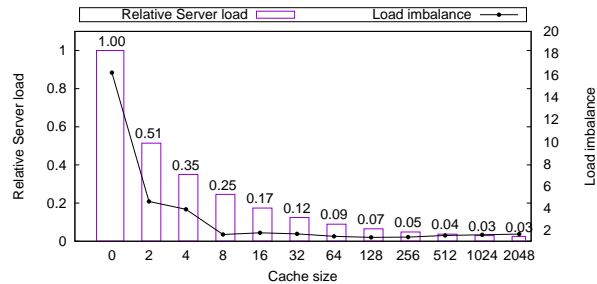}
    \caption{Reduction in relative server load and load-imbalance among caching servers
    as front-end cache size increases.}
        \label{fig:load_imbalance}
\end{figure}

\subsubsection{\textbf{The Need for Cache Resizing:}}
Figure~\ref{fig:load_imbalance} experimentally shows the effect of 
increasing the front-end cache size on both back-end \textit{load-imbalance reduction} 
and decreasing the workload at the back-end. 
In this experiment, 8 memcached shards are deployed to serve back-end lookups and 20 
clients send lookup requests following a significantly skewed Zipfian distribution 
(s = 1.5). The size of the key space is 1 million and the total number of lookups is 
10 millions. The front-end cache size at each client is varied from 0 cachelines (no 
cache) to 2048 cachelines ($\approx$0.2\% of the key space). Front-end caches use 
\algoname's replacement policy and a ratio of 4:1 is maintained between \algoname's
tracker size and \algoname's cache size. We define back-end load-imbalance
as the workload ratio between the most loaded server and the least loaded server.
The target load-imbalance $I_t$ is set to 1.5. As shown in Figure~\ref{fig:load_imbalance},
processing all the lookups from the back-end caching servers (front-end cache size = 0) leads 
to a significant load-imbalance of 16.26 among the caching servers. This means that 
the most loaded caching server receives 16.26 times the number of lookup requests received by
the least loaded caching server. As the front-end cache size increases, the
server size load-imbalance drastically decreases. As shown, a front-end cache
of size 64 cache lines at each client reduces the load-imbalance to 1.44 (an order
of magnitude less load-imbalance across the caching servers) achieving 
the target load-imbalance $I_t = 1.5$. Increasing the front-end cache size beyond
64 cache lines only reduces the back-end aggregated load but not 
the back-end load-imbalance. 
The {\it relative server load} is calculated by comparing the  server load for a given front-end cache size to the server
load when there is no front-end caching (cache size = 0).
Figure~\ref{fig:load_imbalance} demonstrates
the reduction in the relative server load as the front-end cache size increases. 
However, the benefit of doubling the cache
size proportionally decays with the key hotness distribution. As shown in 
Figure~\ref{fig:load_imbalance}, the first 64 cachelines reduce the relative server
load by 91\% while the second 64 cachelines reduce the relative server load by 
only 2\% more.

The failure of the "one size fits all" design strategy suggests
that statically allocating fixed cache and tracker sizes to all front-end 
servers is not ideal. Each front-end server should independently and adaptively
be configured according to the key access distribution it serves. 
Also, changes in workloads 
can alter the key access distribution, the skew level, or the set of hot keys.
For example, social networks and web front-end servers that serve different 
geographical regions might experience different key access distributions and different local 
trends (e.g., \#miami vs. \#ny). Similarly, in large scale data processing pipelines, 
several applications are deployed on top of a shared caching layer. Front-end servers of
different applications serve accesses that might be interested in different partitions
of the data and hence experience different key access distributions and local trends.
Therefore, \algoname's cache resizing algorithm learns the key access distribution 
independently at each front-end and dynamically resizes the cache and the tracker to 
achieve lookup load-imbalance target $I_t$. \algoname is designed to reduce the
front-end cache size that achieves $I_t$. Any increase in the front-end cache size 
beyond \algoname's recommendation mainly decreases back-end load and should 
consider other conflicting parameters such as the additional cost of the memory cost, 
the cost of updates and maintaining the additional cached keys, and the percentage of 
back-end load reduction that results from allocating additional front-end caches.

\subsubsection{\textbf{\algoname: Cache Resizing Algorithm: }}\label{sub:expanding} 
Front-end servers use \algoname to minimize the cache size that achieves a target 
load-imbalance $I_t$. Initially, front-end servers are configured with no front-end caches. 
The system administrator configures \algoname by an input {\em target load-imbalance} 
parameter $I_t$ that determines the maximum tolerable imbalance between the most loaded 
and least loaded back-end caching servers. Afterwards, \algoname expands both tracker and 
cache sizes until the current load-imbalance achieves the inequality $I_c \le I_t$. 

Algorithm~\ref{algo:resizing} describes \algoname's cache resizing algorithm.  
\algoname divides the timeline into epochs and each epoch consists of $E$ accesses. 
Algorithm~\ref{algo:resizing} is executed at the end of each epoch. The epoch size 
$E$ is proportional to the tracker size $K$ and is dynamically updated to guarantee 
that $E \ge K$ (Line~\ref{line:epoch_size}). This inequality is required to guarantee 
that \algoname does not trigger consecutive resizes before the cache and the tracker 
are filled with keys. During each epoch, \algoname tracks the number
of lookups sent to every back-end caching server. In addition, \algoname tracks the total
number of cache hits and tracker hits during this epoch.
At the end of each epoch, 
\algoname calculates the current load-imbalance $I_c$ as the ratio between the highest 
and the lowest load on back-end servers during this epoch.  Also, \algoname calculates the
current average hit per cached key $\alpha_c$. $\alpha_c$ equals the total cache hits in
the current epoch divided by the cache size. Similarly, \algoname calculates the current average 
hit per tracked but not cache key $\alpha_{k-c}$. 
\algoname compares $I_c$ to $I_t$ and decides on a resizing action as follows.

\begin{enumerate}
\item $I_c > I_t$ (Line~\ref{line:imbalanced}), this means that the target load-imbalance is not 
achieved. \algoname follows the binary search algorithm in searching
for the front-end cache size that achieves $I_t$. Therefore, \algoname decides to 
double the front-end cache size (Line~\ref{line:double_cache}). 
As a result, \algoname doubles the tracker size as well to
maintain a tracker to cache size ratio of at least 2, $K \ge 2\cdot C$ 
(Line~\ref{line:double_tracker}).
In addition, \algoname uses a local variable $\alpha_t$ to capture the quality 
of the cached keys when $I_t$ is first achieved. Initially, $\alpha_t = 0$. \algoname then sets $\alpha_t$ to the 
average hits per cache-line $\alpha_c$ during the current epoch (Line~\ref{line:alpha}).  
In subsequent epochs, $\alpha_t$ is used to detect changes in workload. 

\item $I_c \le I_t$ (Line~\ref{line:balanced}), this means that the target load-imbalance has 
been achieved. However, changes in workload could alter the quality of the cached keys. Therefore, 
\algoname uses $\alpha_t$ to detect and handle changes in workload in future epochs as 
explained below. 

\end{enumerate}
\begin{algorithm}[!h]
\caption{\algoname's elastic resizing algorithm.}
 \label{algo:resizing}
\begin{flushleft}
\textbf{State:} $S_c$: keys in the cache, $S_k$: keys in the tracker, C: cache capacity, K: tracker capacity, $\alpha_c$: average hits per key in $S_c$ in the current epoch, $\alpha_{k-c}$: average hits per key in $S_{k-c}$ in the current epoch, $I_c$: current load-imbalance,
and $\alpha_t$: target average hit per key\newline
\textbf{Input:} $I_t$
\end{flushleft}
\begin{algorithmic}[1]
	\If {$I_c > I_t$} \label{line:imbalanced}
		\State resize($S_c$, $2 \times C$) \label{line:double_cache}
        \State resize($S_k$, $2\times K$) \label{line:double_tracker}
        \State E := max (E, $K$) \label{line:epoch_size}
        \State Let $\alpha_t$ = $\alpha_c$ \label{line:alpha}
  	\Else  \label{line:balanced}
	\If {$\alpha_c < (1-\epsilon).\alpha_t$  and $\alpha_{k-c} < (1-\epsilon).\alpha_t$ } \label{line:case1}
		\State resize($S_c$, $\frac{C}{2}$) \label{line:half_cache}
        \State resize($S_k$, $\frac{K}{2}$) \label{line:half_tracker}
    \ElsIf{$\alpha_c < (1-\epsilon).\alpha_t$  and $\alpha_{k-c} > (1-\epsilon).\alpha_t$}\label{line:case2}
        \State half\_life\_time\_decay() \label{line:hlt}
    \Else
        \State do\_nothing() \label{line:do_nothing}
    \EndIf
  \EndIf
\end{algorithmic}
\end{algorithm}

$\alpha_t$  is reset whenever the inequality $I_c \le I_t$ is violated and Algorithm~\ref{algo:resizing} expands cache and 
tracker sizes. Ideally, when the inequality $I_c \le I_t$ holds, 
keys in the cache (the set $S_c$) achieve $\alpha_t$ hits per cache-line during 
every epoch while keys in the tracker but not in the cache (the set $S_{k-c}$) do
not achieve $\alpha_t$. This happens because keys in the set $S_{k-c}$ are less
hot than keys in the set $S_c$. $\alpha_t$ represents a target hit-rate per cache-line
for future epochs. Therefore, if keys in the cache do not meet the target $\alpha_t$
in a following epoch, this indicates that the quality of the cached keys has changed and
an action needs to be taken as follows.

\begin{enumerate}
\item Case 1: keys in $S_c$, on the average,
do not achieve $\alpha_t$ hits per cacheline and keys in $S_{k-c}$ do not 
achieve $\alpha_t$ hits as well (Line~\ref{line:case1}). This indicates that the quality 
of the cached keys decreased. In response. \algoname shrinks both the cache
and the tracker sizes (Lines~\ref{line:half_cache} and~\ref{line:half_tracker}).
If shrinking both cache and tracker sizes results in a violation of the inequality 
$I_c < I_t$, Algorithm~\ref{algo:resizing} doubles both tracker and cache sizes in
the following epoch and $\alpha_t$ is reset as a result. 
In Line~\ref{line:case1}, we compare the average hits per
key in both $S_c$ and $S_{k-c}$ to $(1-\epsilon)\cdot \alpha_t$ instead of $\alpha_t$. Note that $\epsilon$ is a small constant $<<<1$ that is used to avoid unnecessary 
resizing actions due to insignificant statistical variations.

\item Case 2: keys in $S_c$ do not achieve $\alpha_t$ while
keys in $S_{k-c}$ achieve $\alpha_t$ (Line~\ref{line:case2}). This signals that
the set of hot keys is changing and keys in $S_{k-c}$ are becoming hotter
than keys in $S_c$. For this, \algoname triggers a half-life time decaying 
algorithm that halves the hotness of all cached and tracked 
keys (Line~\ref{line:hlt}). 
This decaying algorithm aims to forget old trends that are no longer 
hot to be cached (e.g., Gangnam style song). Different decaying algorithms have 
been developed in the 
literature~\cite{cormode2008exponentially,cormode2009forward,cohen2003maintaining}.
Therefore, this paper only focuses on the resizing algorithm details without
implementing a decaying algorithm.

\item Case 3: keys in $S_c$ achieve $\alpha_t$ while
keys in $S_{k-c}$ do not achieve $\alpha_t$. This means that the quality of
the cached keys has not changed and therefore, \algoname does not take any
action. Similarly, if keys in both sets $S_c$ and $S_{k-c}$ achieve $\alpha_t$, 
\algoname does not take any action as long as the inequality $I_c < I_t$ holds 
(Line~\ref{line:do_nothing}).

\end{enumerate}

\section{Experimental Evaluation} \label{sec:evaluation}

In this section, we evaluate \algoname's caching algorithm and \algoname's adaptive 
resizing algorithm. We choose to compare \algoname to traditional and widely 
used replacement policies like LRU and LFU. In addition, we compare \algoname to 
both ARC~\cite{megiddo2003arc} and LRU-k~\cite{o1993lruk}. As stated 
in~\cite{megiddo2003arc}, ARC, in its online auto-configuration setting, achieves 
comparable performance to LRU-2 (which is the most responsive LRU-k )~\cite{o1993lruk, o1999lruk2},
2Q~\cite{johnson1994x3}, LRFU~\cite{lee2001lrfu}, and LIRS~\cite{jiang2002lirs} 
even when these policies are perfectly tuned offline. Also, ARC outperforms the online
adaptive replacement policy MQ~\cite{zhou2001multi}. Therefore, we compare with ARC and LRU-2 as representatives of these different polices. 
The experimental setup is explained in 
Section~\ref{sub:experiment_setup}. First, we compare the hit rates of \algoname's 
cache algorithm to LRU, LFU, ARC, and LRU-2 hit rates for different front-end cache 
sizes in Section~\ref{sub:hit_rate}. Then, we compare the required
front-end cache size for each replacement policy to achieve a target back-end 
load-imbalance $I_t$ in Section~\ref{sub:backend-load-imbalance}. In 
Section~\ref{sub:end-to-end}, we provide an end-to-end evaluation of front-end caches 
comparing the end-to-end performance of \algoname, LRU, LFU, ARC, and LRU-2 on different workloads with the configuration where no front-end cache is deployed. Finally,
\algoname's resizing algorithm is evaluated in Section~\ref{sub:adaptive_resizing}.

\subsection{Experiment Setup} \label{sub:experiment_setup}
We deploy 8 instances of memcached~\cite{memcached} on a small cluster of 
4 caching servers (2 memcached instance per server). 
Each caching server has an Intel(R) Xeon(R) CPU E31235
with 4GB RAM dedicated to each memcached instance.

Dedicated client machines are used to generate client workloads. Each
client machine executes multiple client threads to submit workloads to
caching servers. Client threads use Spymemcached 2.11.4~\cite{spymemcached},
a Java-based memcached client, to communicate with memcached cluster.
Spymemcached provides communication abstractions that distribute workload
among caching servers using 
\textit{consistent hashing}~\cite{karger1997consistent}. We slightly modified
Spymemcached to monitor the workload per back-end server at each front-end. Client 
threads use Yahoo! Cloud Serving Benchmark (YCSB)~\cite{cooper2010benchmarking} to 
generate workloads for the experiments. YCSB is a standard key/value store
benchmarking framework. YCSB is used to generate key/value store requests
such as \textit{Get}, \textit{Set}, and \textit{Insert}. YCSB enables 
configuring the ratio between read (Get) and write (Set) accesses. Also,
YCSB allows the generation of accesses that follow different access distributions.
As YCSB is CPU-intensive, client machines run at most 20 client threads per
machine to avoid contention among client threads. During our experiments,
we realized that YCSB's ScrambledZipfian workload generator has a bug as it 
generates Zipfian workload distributions with significantly less skew than the skew 
level it is configured with. Therefore, we use YCSB's ZipfianGenerator instead of
YCSB's ScrambledZipfian.

Our experiments use different variations of YCSB core workloads.
Workloads consist of 1 million key/value pairs. Each key consists
of a common prefix \textit{"usertable:"} and a unique ID. We use a value size of 
750 KB making a dataset of size 715GB. Experiments use read intensive 
workloads that follow Tao's~\cite{bronson2013tao} read-to-write ratio
of 99.8\% reads and 0.2\% updates. Unless otherwise specified, 
experiments consist of 10 million key accesses sampled from different 
access distributions such as Zipfian (s = 0.90, 0.99, or 1.2)
and uniform. Client threads submit access requests back-to-back.
Each client thread can have only one outgoing request. Clients submit a 
new request as soon as they receive an acknowledgement for their 
outgoing request.

\subsection{Hit Rate} \label{sub:hit_rate}

\begin{figure*}[ht!]
    \centering
    \begin{subfigure}[t]{0.32\textwidth}
        \includegraphics[width=\columnwidth]{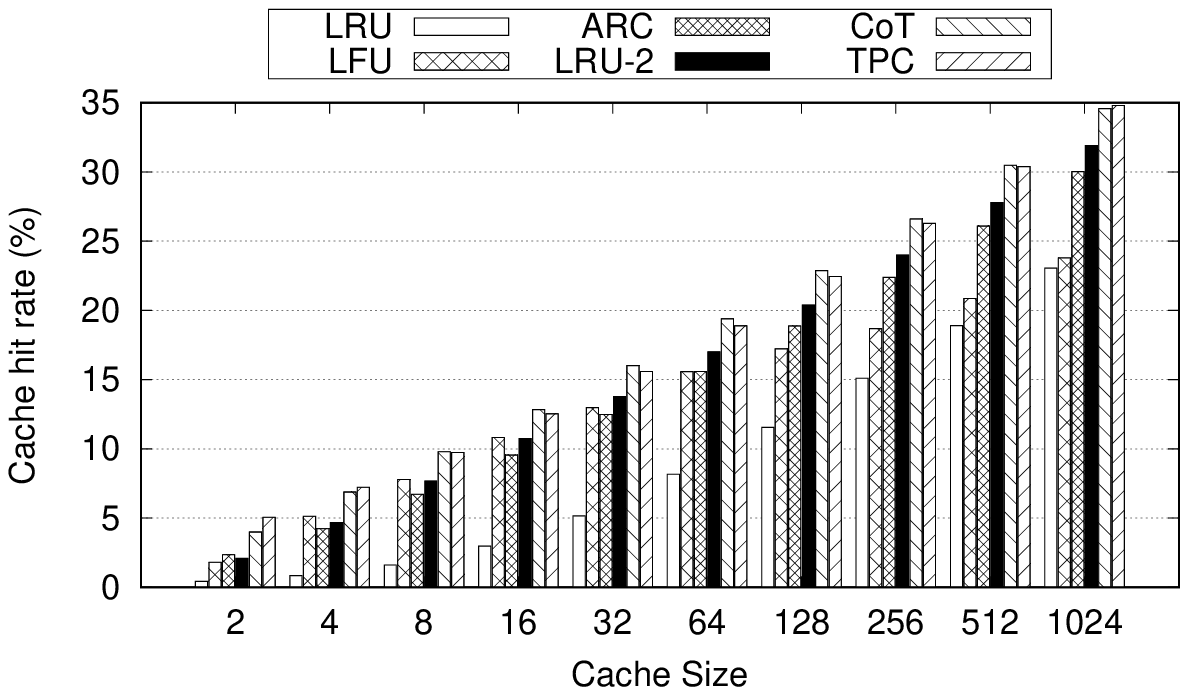}
        \caption{Zipfian 0.90}
        \label{fig:hitrate_zipf90}
    \end{subfigure}
    \begin{subfigure}[t]{0.32\textwidth}
        \includegraphics[width=\columnwidth]{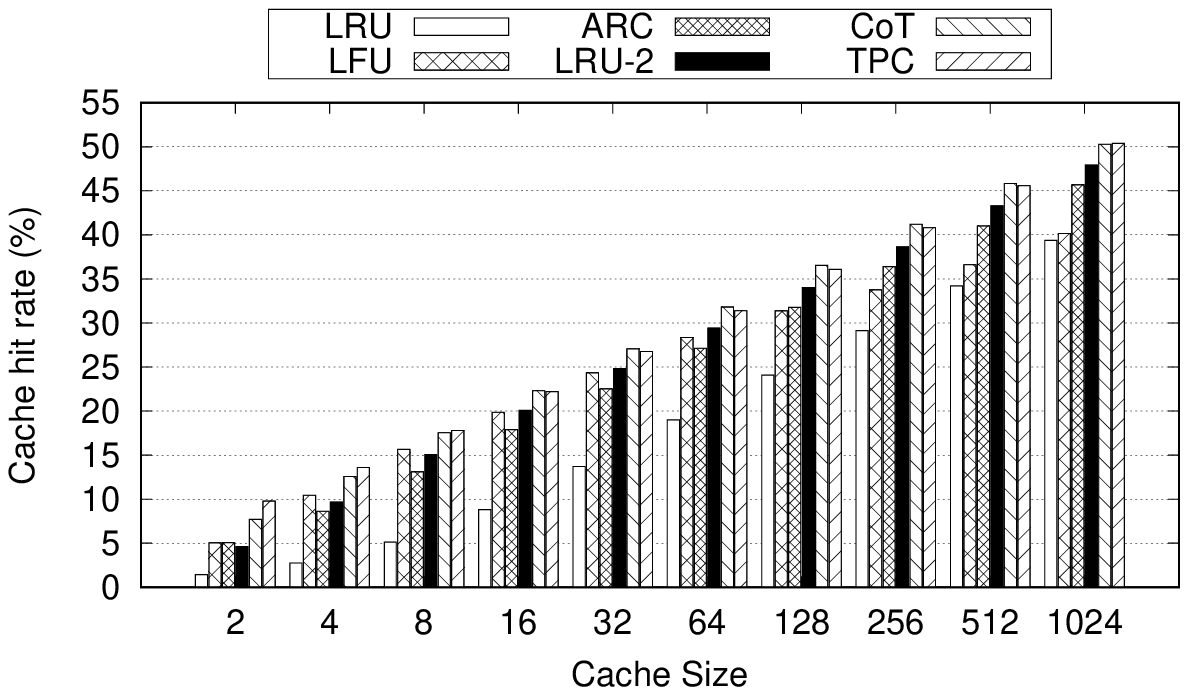}
        \caption{Zipfian 0.99}
        \label{fig:hitrate_zipf99}
    \end{subfigure}
    \begin{subfigure}[t]{0.32\textwidth}
        \includegraphics[width=\columnwidth]{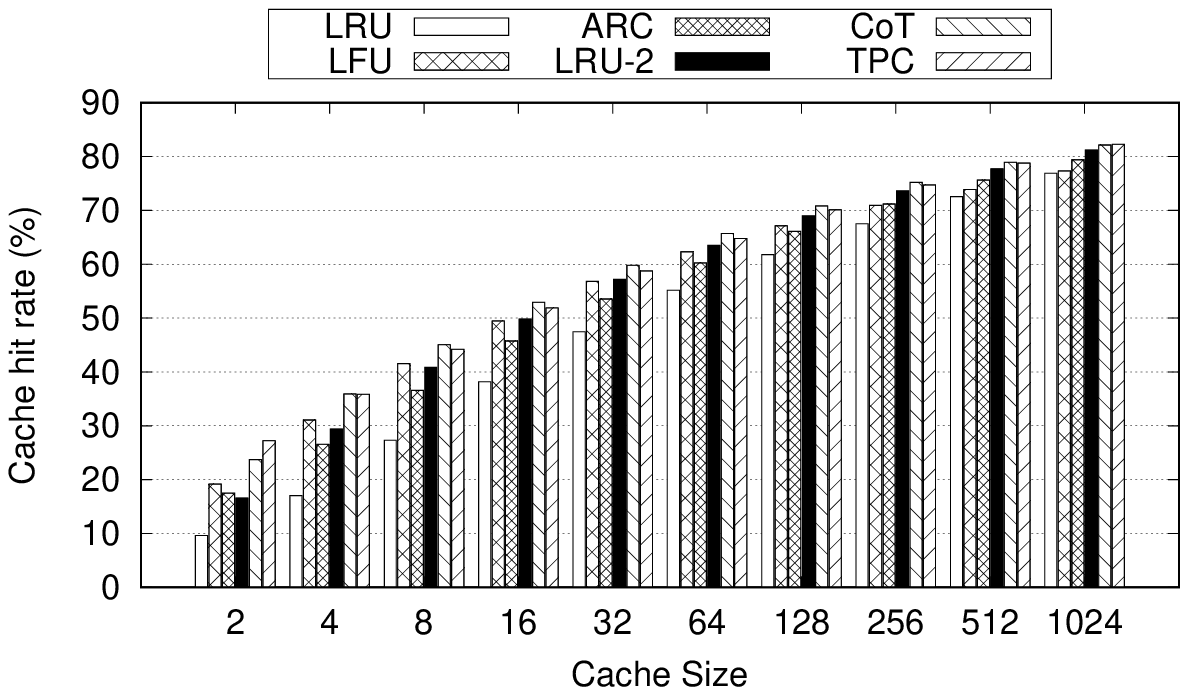}
        \caption{Zipfian 1.20}
        \label{fig:hitrate_zipf120}
    \end{subfigure} 
    \caption{Comparison of LRU, LFU, ARC, LRU-2, \algoname and TPC's hit rates using 
    Zipfian access distribution with different skew parameter values (s= 0.90, 0.99, 1.20) }
    \label{fig:hitrate}
\end{figure*}

The first experiment compares \algoname's hit rate to LRU, LFU, ARC, and LRU-2
hit rates using equal cache sizes for all replacement policies. 20 client threads are provisioned
on one client machine and each cache client maintains its own cache.
The cache size is varied from a very small cache of 2 cache-lines to 1024
cache-lines. The hit rate is compared using different
Zipfian access distributions with skew parameter values s = 0.90, 0.99, 
and 1.2 as shown in Figures~\ref{fig:hitrate_zipf90},~\ref{fig:hitrate_zipf99},
and~\ref{fig:hitrate_zipf120} respectively. \algoname's
tracker to cache size ratio determines how many tracking nodes
are used for every cache-line. 
\algoname automatically detects the ideal
tracker to cache ratio for any workload by fixing the cache size and doubling the 
tracker size until the observed hit-rate gains from increasing the tracker size 
are insignificant i.e., the observed hit-rate saturates.
The tracker to 
cache size ratio decreases as the workload skew increases. A workload
with high skew simplifies the task of distinguishing hot keys from cold 
keys and hence, \algoname requires a smaller tracker size to 
successfully filter hot keys from cold keys.  Note that LRU-2 is also configured
with the same history to cache size as \algoname's tracker to cache size.
In this experiment, for each skew level, 
\algoname's tracker to cache size ratio is varied as follows:
16:1 for Zipfian 0.9, 8:1 for Zipfian 0.99, and 4:1 for Zipfian 1.2.
Note that \algoname's tracker maintains only the meta-data of tracked keys. Each tracker 
node consists of a read counter and a write counter with 8 bytes of memory overhead per 
tracking node. In real-world workloads, value sizes vary from few hundreds KBs to few MBs. 
For example, Google's Bigtable~\cite{chang2008bigtable} uses a value size of 64 MB.
Therefore, a memory overhead of at most $\frac{1}{8}$ KB (16 tracker nodes * 8 bytes)
per cache-line is negligible.

In Figures~\ref{fig:hitrate}, the \textit{x-axis}
represents the cache size expressed as the number of cache-lines. The 
\textit{y-axis} represents the front-end cache hit rate (\%) as a percentage of the 
total workload size. At each cache size, 
the cache hit rates are reported for LRU, LFU, ARC, LRU-2, and \algoname cache 
replacement policies. In addition, TPC represents the theoretically calculated hit-rate
from the Zipfian distribution CDF if a perfect cache with the same cache size is deployed. For
example, a perfect cache of size 2 cache-lines stores the hot most 2 keys and 
hence any access to these 2 keys results in a cache hit while accesses to other keys 
result in cache misses.

As shown in Figure~\ref{fig:hitrate_zipf90}, \algoname surpasses LRU, LFU, ARC, and LRU-2
hit rates at all cache sizes. In fact, \algoname achieves almost similar hit-rate to
the TPC hit-rate. In Figure~\ref{fig:hitrate_zipf90}, \algoname outperforms TPC for some
cache size which is counter intuitive. This happens as TPC is theoretically calculated 
using the Zipfian CDF while \algoname's hit-rate is calculate out of YCSB's sampled
distributions which are approximate distributions. In addition, \algoname achieves higher 
hit-rates than both LRU and LFU with \textbf{75\% less cache-lines}. As shown, \algoname 
with 512 cache-lines achieves 10\% more hits than both LRU and LFU with 2048 
cache-lines. Also,
\algoname achieves higher hit rate than ARC using \textbf{50\% less cache-lines}.
In fact, \algoname configured with 512 cache-lines achieves 2\% more hits than ARC 
with 1024 cache-lines. Taking tracking memory overhead into account, 
\algoname maintains a tracker to cache size ratio of 16:1 for this workload (Zipfian 0.9).
This means that \algoname adds an overhead of 128 bytes (16 tracking nodes * 8 bytes each) 
per cache-line. The percentage of \algoname's tracking memory overhead decreases as the 
cache-line size increases. For example, \algoname introduces a tracking overhead of 
0.02\% when the cache-line size is 750KB. Finally, \algoname consistently achieves 8-10\%
higher hit-rate than LRU-2 configured with the same history and cache sizes as \algoname's
tracker and cache sizes.

Similarly, as illustrated in Figures~\ref{fig:hitrate_zipf99} 
and~\ref{fig:hitrate_zipf120}, \algoname outpaces LRU,
LFU, ARC, and LRU-2 hit rates at all different cache sizes. Figure~\ref{fig:hitrate_zipf99}
shows that a configuration of \algoname using 512 cache-lines achieves 3\% more hits than
both configurations of LRU and LFU with 2048 cache-lines. Also, \algoname consistently 
outperforms ARC's hit rate with 50\% less cache-lines. Finally, \algoname achieves 3-7\%
higher hit-rate than LRU-2 configured with the same history and cache sizes. 
Figures~\ref{fig:hitrate_zipf99} and~\ref{fig:hitrate_zipf120} highlight that increasing 
workload skew decreases the advantage of \algoname. As workload skew increases,
the ability of LRU, LFU, ARC, LRU-2 to distinguish between hot and cold keys 
increases and hence \algoname's preeminence decreases.

\subsection{Back-End Load-Imbalance} \label{sub:backend-load-imbalance}
In this section, we compare the required front-end cache sizes for different 
replacement policies to achieve a back-end load-imbalance target $I_t$. Different 
skewed workloads are used, namely, Zipfian
s = 0.9, s = 0.99, and s = 1.2. For each distribution, we first measure
the back-end load-imbalance when no front-end cache is used. A back-end 
load-imbalance target $I_t$ is set to $I_t = 1.1$. This means that the 
back-end is load balanced if the most loaded back-end
server processes at most 10\% more lookups than the least loaded back-end server.
We evaluate the back-end load-imbalance while increasing the front-end cache
size using different cache replacement policies, namely, LRU, LFU, ARC, LRU-2,
and \algoname. In this experiment, \algoname uses the same tracker-to-cache
size ratio as in Section~\ref{sub:hit_rate}. For each replacement policy, 
we report the minimum required number of cache-lines to achieve $I_t$.

\begin{table*}[]

\centering
\begin{tabular}{ | l| l | l | l| l| l| l|}
 \hline
 \multirow{2}{*}{Dist.} &
 \multirow{2}{*}{\begin{tabular}[t]{@{}l@{}}
Load-\\imbalance \\ No cache\end{tabular}}& \multicolumn{5}{l|}{\begin{tabular}[t]{@{}l@{}}
Number of cache-lines \\ to achieve $I_t = 1.1$ \end{tabular}}  \\ \cline{3-7}
& & LRU & LFU & ARC & LRU-2 & \algoname \\ \hline
Zipf 0.9  & 1.35 &  64 & 16 & 16 & 8 & 8 \\ \hline
Zipf 0.99  & 1.73 &  128 & 16 & 16 & 16 & 8 \\ \hline
Zipf 1.20  & 4.18 &  2048 & 2048 & 1024 & 1024 & 512 \\ \hline
\end{tabular}
\caption{The minimum required number of cache-lines for
different replacement policies to achieve a back-end load-imbalance
target $I_t = 1.1$ for different workload distributions.}
\label{table:imbalance}
\end{table*}

Table~\ref{table:imbalance} summarizes the reported results for
different distributions using LRU, LFU, ARC, LRU-2, and \algoname replacement policies.
For each distribution, the initial back-end load-imbalance is 
measured using no front-end cache. As shown, the initial load-imbalances
for Zipf 0.9, Zipf 0.99, and Zipf 1.20 are 1.35, 1.73, and 4.18 respectively.
For each distribution, the minimum required number of cache-lines for
LRU, LFU, ARC, and \algoname to achieve a target load-imbalance of $I_t = 1.1$
is reported. As shown, \algoname requires \textbf{50\% to 93.75\% less cache-lines}
than other replacement policies to achieve $I_t$. Since LRU-2 is configured with a
history size equals to \algoname's tracker size, LRU-2 requires the second least number
of cache-lines to achieve $I_t$.

\subsection{End-to-End Evaluation} \label{sub:end-to-end}

\begin{figure}[ht!]
    \includegraphics[width=\columnwidth]{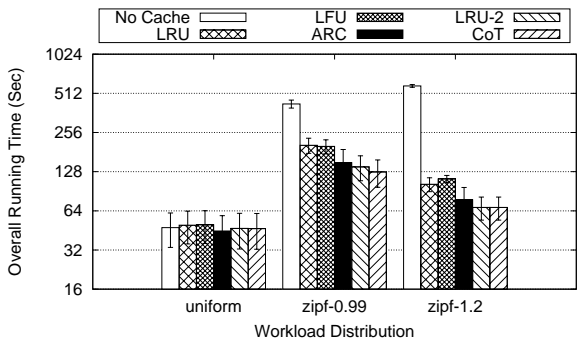}
    \caption{The effect of front-end caching on the end-to-end 
    overall running time of 1M lookups using different workload distributions.}
    \label{fig:end-to-end}
\end{figure}

In this section, we evaluate the effect of front-end caches using LRU,
LFU, ARC, LRU-2, and \algoname replacement policies on the overall running time
of different workloads. This experiment also demonstrates the overhead of 
front-end caches on the overall running time. In this experiment, we use
3 different workload distributions, namely, uniform, Zipfian (s = 0.99), and 
Zipfian (s = 1.2) distributions as shown in Figure~\ref{fig:end-to-end}. For 
all the three workloads, each replacement policy 
is configured with 512 cache-lines. Also, \algoname and LRU-2 maintains a tracker (history)
to cache size ratio of 8:1 for Zipfian 0.99 and 4:1 for both Zipfian 1.2 and uniform 
distributions. In this experiment, a total of 1M accesses are sent to the caching servers
by 20 client threads running on one client machine.
Each experiment is executed 10 times and the average
overall running time with 95\% confidence intervals are reported in 
Figure~\ref{fig:end-to-end}.

In this experiment, the front-end servers are
allocated in the same cluster as the back-end servers. The average 
Round-Trip Time (RTT) between front-end machines and back-end machines
is 244$\mu$s. This small RTT allows us to fairly measure the overhead 
of front-end caches by minimizing the performance advantages achieved by front-end
cache hits. In real-world deployments where front-end servers are deployed in
edge-datacenters and the RTT between front-end servers and back-end servers
is in order of \textbf{10s of ms}, front-end caches achieve more
significant performance gains.

The uniform workload is used to measure the overhead of front-end caches.
In a uniform workload, all keys in the key space are equally hot and
front-end caches cannot take any advantage of workload skew to benefit some 
keys over others. Therefore, front-end caches only introduce the overhead
of maintaining the cache without achieving any significant performance gains. 
As shown in Figure~\ref{fig:end-to-end}, there is no significant statistical 
difference between the overall running time when there is no front-end cache 
and when there is a small front-end cache with different replacement policies.
Adding a small front-end cache does not incur running time overhead even for
replacement policies that use a heap (e.g., LFU, LRU-2, and \algoname).

The workloads Zipfian 0.99 and Zipfian 1.2 are used to show the advantage of
front-end caches even when the network delays between front-end servers
and back-end servers are minimal. As shown in Figure~\ref{fig:end-to-end},
workload skew results in significant overall running time overhead in the absence
of front-end caches. This happens
because the most loaded server introduces a performance bottleneck especially under
thrashing (managing 20 connections, one from each client thread). As the load-imbalance
increases, the effect of this bottleneck is worsen. Specifically, in 
Figure~\ref{fig:end-to-end}, the overall running time of Zipfian 0.99 and Zipfian 1.2 
workloads are respectively 8.9x and 12.27x of the uniform workload when no front-end cache 
is deployed. Deploying a small front-end cache of 512 cachelines significantly reduces
the effect of back-end bottlenecks. Deploying a \algoname small cache in the front-end
results in 70\% running time reduction for Zipfian 0.99 and 88\% running time reduction
for Zipfian 1.2 in comparison to having no front-end cache. Other replacement policies
achieve running time reductions of 52\% to 67\% for Zipfian 0.99 and 80\% to 88\% for 
Zipfian 1.2. LRU-2 achieves the second best average overall running time 
after \algoname with no significant statistical difference between the two policies. Since both policies use the same tracker (history) size, this again suggests that having a bigger tracker helps separate cold and noisy
keys from hot keys. Since the ideal tracker to cache size ratio differs from
one workload to another, having an automatic and dynamic way to configure 
this ratio at run-time while serving workload gives \algoname a big leap
over statically configured replacement policies.

\begin{figure}[ht!]
    \includegraphics[width=\columnwidth]{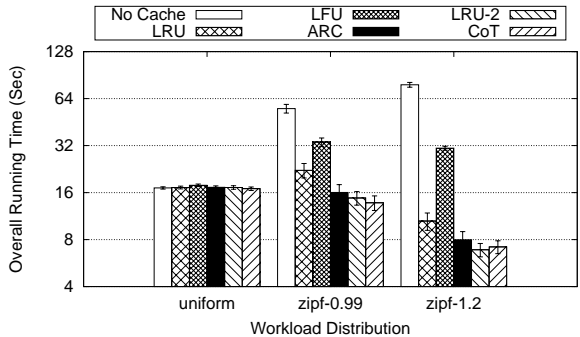}
    \caption{The effect of front-end caching on the end-to-end 
    overall running time of 50K lookups using different workload distributions sent by only \textbf{one} client thread.}
    \label{fig:end-to-end-50K}
\end{figure}

To isolate the effect of both front-end and back-end thrashing on
the overall running time, we run the same experiment with only one client
thread that executes 50K lookups (1M/20) and we report
the results of this experiment in Figure~\ref{fig:end-to-end-50K}. The first
interesting observation of this experiment is that the overall running time 
of Zipfian 0.99 and Zipfian 1.2 workloads are respectively 3.2x and 4.5x of 
the uniform workload when no front-end cache is deployed. These numbers
are proportional to the load-imbalance factors of these two distributions
(1.73 for Zipfian 0.99 and 4.18 for Zipfian 1.2). These factors are significantly worsen under thrashing as shown in the previous experiment.
The second
interesting observation is that deploying a small front-end cache in a non-thrashing environment results in a lower overall running time
for skewed workload (e.g., Zipfian 0.99 and Zipfian 1.2) than for a uniform
workload. This occurs because front-end caches eliminate back-end load-imbalance and locally serve lookups as well.

\subsection{Adaptive Resizing}\label{sub:adaptive_resizing}

\begin{figure}[ht!]
    \centering
    \begin{subfigure}[t]{\columnwidth}
	\includegraphics[width=\columnwidth]{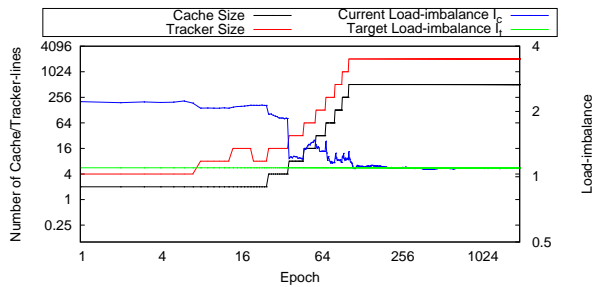}        		\caption{Changes in cache and tracker sizes and the current load-imbalance $I_c$
	over epochs.}
        \label{fig:zipf120_expand-load}
    \end{subfigure}
    \begin{subfigure}[t]{\columnwidth}
	\includegraphics[width=\columnwidth]{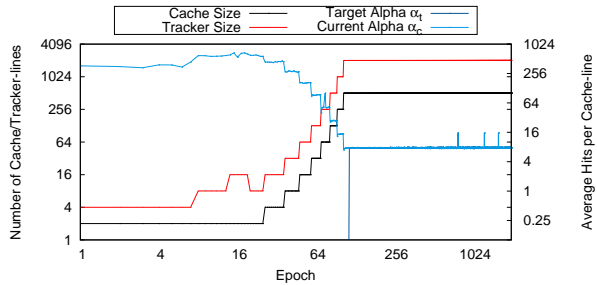}        		\caption{Changes in cache and tracker sizes and the current hit rate per cacheline $\alpha_c$ 
	over epochs.}
        \label{fig:zipf120_expand-alpha}
    \end{subfigure}
    \caption{\algoname adaptively expands tracker and cache sizes to achieve a target load-imbalance 
    $I_t = 1.1$ for a Zipfian 1.2 workload.}
    \label{fig:zipf120_expand}
\end{figure}

\begin{figure}[ht!]
    \centering
    \begin{subfigure}[t]{\columnwidth}
	\includegraphics[width=\columnwidth]{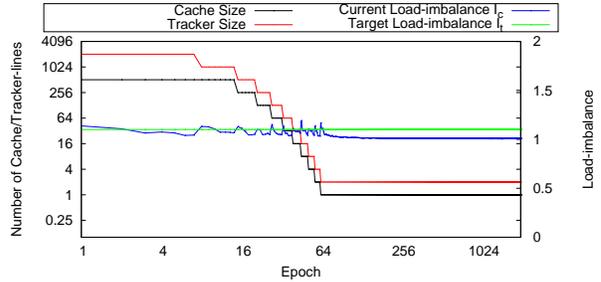}        		\caption{Changes in cache and tracker sizes and the current load-imbalance $I_c$
	over epochs.}
        \label{fig:zipf120_shrink-load}
    \end{subfigure}
    \begin{subfigure}[t]{\columnwidth}
	\includegraphics[width=\columnwidth]{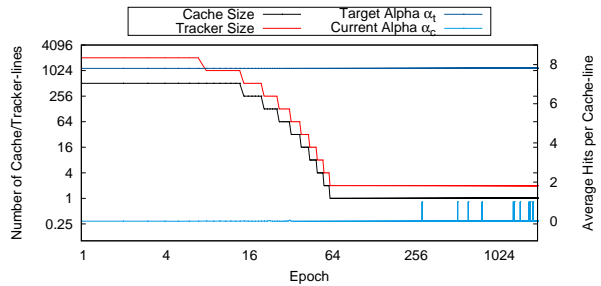}        		\caption{Changes in cache and tracker sizes and the current hit rate per cache-line $\alpha_c$ 
	over epochs.}
        \label{fig:zipf120_shrink-alpha}
    \end{subfigure}
    \caption{\algoname adaptively shrinks tracker and cache sizes in response to changing the workload to uniform.}
    \label{fig:zipf120_shrink}
\end{figure}

This section evaluates \algoname's auto-configure and resizing algorithms.
\textit{First}, we configure a front-end client that serves a Zipfian 1.2 
workload with a tiny cache of size \textit{two} cachelines and a tracker of 
size of \textit{four} tracking entries. This experiment aims to show
how \algoname expands cache and tracker sizes to achieve
a target load-imbalance $I_t$ as shown in Figure~\ref{fig:zipf120_expand}. 
After \algoname reaches the cache size that achieves $I_t$, the average hit 
per cache-line $\alpha_t$ is recorded as explained in 
Algorithm~\ref{algo:resizing}. \textit{Second}, we alter the workload 
distribution to uniform and monitors how \algoname 
shrinks tracker and cache sizes in response to workload changes
without violating the load-imbalance target $I_t$ in 
Figure~\ref{fig:zipf120_shrink}. In both experiments, $I_t$ is set to 1.1 and the epoch size is 5000 accesses. In both Figures~\ref{fig:zipf120_expand-load} and~\ref{fig:zipf120_shrink-load},
the x-axis represents the epoch number, the left y-axis represents the number of tracker and cache lines, and the right y-axis represents the load-imbalance. The black and red lines represent cache and tracker sizes
respectively with respect to the left y-axis. The blue and green lines
represent the current load-imbalance and the target load-imbalance respectively with respect to the right y-axis. Same axis description applies for both 
Figures~\ref{fig:zipf120_expand-alpha} and~\ref{fig:zipf120_shrink-alpha} except
that the right y-axis represents the average hit per cache-line during each epoch.
Also, the light blue and the dark blue lines represent the current average hit
per cache-line and the target hit per cache-line at each epoch with respect to the
right y-axis.

In Figure~\ref{fig:zipf120_expand-load}, \algoname is initially configured with
a cache of size 2 and a tracker of size 4. \algoname's resizing algorithm 
runs in 2 phases. In the first phase, \algoname discovers the ideal 
tracker-to-cache size ratio that maximizes the hit rate for a fixed cache size for the current 
workload. For this, \algoname fixes the cache size and doubles the tracker size
until doubling the tracker size achieves no significant benefit on the hit rate. 
This is shown in Figure~\ref{fig:zipf120_expand-alpha} in the first 15 epochs. \algoname
allows a warm up period of 5 epochs after each tracker or cache resizing decision.
Notice that increasing the tracker size while fixing the cache size reduces 
the current load-imbalance $I_c$ (shown in Figure~\ref{fig:zipf120_expand-load}) and 
increases the current observed hit per cache-line $\alpha_c$ 
(shown in Figure~\ref{fig:zipf120_expand-alpha}).
Figure~\ref{fig:zipf120_expand-alpha} shows that \algoname first expands the tracker size
to 16 and during the warm up epochs (epochs 10-15), \algoname observes no significant
benefit in terms of $\alpha_c$ when compared to a tracker size of 8. In response, 
\algoname therefore shrinks the tracker size to 8 as shown in the dip in the red line in 
Figure~\ref{fig:zipf120_expand-alpha} at epoch 16. Afterwards, \algoname starts phase 2
searching for the smallest cache size that achieves $I_t$. For this, \algoname
doubles the tracker and caches sizes until the target load-imbalance is achieved
and the inequality $I_c \le I_t$ holds as shown in Figure~\ref{fig:zipf120_expand-load}.
\algoname captures $\alpha_t$ when $I_t$ is first achieved. $\alpha_t$ determines the quality
of the cached keys when $I_t$ is reached for the first time.
In this experiment, \algoname does not trigger resizing if $I_c$ is within 2\% of $I_t$.
Also, as the cache size increases, $\alpha_c$ decreases as the skew of the additionally
cached keys decreases. For a Zipfian 1.2 workload and to achieve $I_t=1.1$, \algoname requires
512 cache-lines and 2048 tracker lines and achieves an average hit per cache-line of $\alpha_t 
= 7.8$ per epoch.

Figure~\ref{fig:zipf120_shrink} shows how \algoname successfully shrinks tracker 
and cache sizes in response to workload skew drop without violating $I_t$. After running the experiment in Figure~\ref{fig:zipf120_expand}, we
alter the workload to uniform. Therefore, \algoname detects a drop in the 
current average hit per cache-line as shown in Figure~\ref{fig:zipf120_shrink-alpha}.
At the same time, \algoname observe that the current load-imbalance $I_c$
achieves the inequality $I_c \le I_t = 1.1$. Therefore, \algoname decides
to shrink both the tracker and cache sizes until either $\alpha_c \approx 
\alpha_t = 7.8$ or $I_t$ is violated or until cache and tracker sizes are
negligible. First, \algoname resets the tracker to cache size ratio to 2:1 and
then searches for the right tracker to cache size ratio for the current workload. Since the workload is uniform, expanding the tracker size beyond 
double the cache size achieves no hit-rate gains as shown in 
Figure~\ref{fig:zipf120_shrink-alpha}. Therefore, \algoname moves to the second phase of shrinking both tracker and cache sizes as long $\alpha_t$
is not achieved and $I_t$ is not violated. As shown, in Figure~\ref{fig:zipf120_shrink}, \algoname shrinks both the tracker and the cache sizes until front-end cache size becomes negligible. As shown in
Figure~\ref{fig:zipf120_shrink-load}, \algoname shrinks cache and tracker sizes while ensuring that the target load-imbalance is not violated.

\section{Related Work} \label{sec:related_work}

Distributed caches are widely deployed to serve social networks and the 
web at scale~\cite{bronson2013tao,nishtala2013scaling,zakhary2017caching}.
Real-world workloads are typically skewed with few keys that are 
significantly hotter than other keys~\cite{huang2014characterizing}.
This skew can cause load-imbalance among the caching servers.
Load-imbalancing negatively affects the overall performance of the caching
layer. Therefore, many works in the literature have 
addressed the load-imbalacing problem from different angles. Solutions use
different load-monitoring techniques (e.g., centralized 
tracking~\cite{adya2016slicer,hwang2013adaptive,adya2010centrifuge,wu2019autoscaling}, server-side 
tracking~\cite{hong2013understanding,cheng2015memory}, and client-side 
tracking~\cite{fan2011small,jin2017netcache}). Based on the load-monitoring, different 
solutions redistribute keys among caching servers at 
different granularities. The following paragraphs summarize the related works 
under different categories.

\textbf{Centralized load-monitoring:} Slicer~\cite{adya2016slicer} separates the 
data serving plane from the control plane. The key space is divided into 
\textit{slices} where each slice is assigned to one or more servers. The control 
plane is a centralized system component that collects the access information of
each slice and the workload per server. The control plane periodically runs 
an optimization that generates a new slice assignment. This assignment might
result in redistributing, repartitioning, or replicating slices among servers to 
achieve better load-balancing. Unlike in Centrifuge~\cite{adya2010centrifuge}, 
Slicer does not use consistent hashing to map keys to servers. Instead, Slicer 
distributes the generated assignments to the front-end servers to allow them to 
locate keys. Also, Slicer highly replicates the centralized control plane
to achieve high availability and to solve the fault-tolerance problem in both 
Centrifuge~\cite{adya2010centrifuge} and in~\cite{cheng2015memory}. \algoname is complementary
to systems like Slicer. Our goal is to cache heavy hitters at front-end servers
to reduce key skew at back-end caching servers and hence, reduce Slicer's initiated
re-configurations. Our focus is on developing a 
replacement policy and an adaptive cache resizing algorithm to enhance 
the performance of front-end caches. Also, our approach is distributed 
and front-end driven that does not require any system component to develop
a global view of the workload. This allows \algoname to scale to thousands of
front-end servers without introducing any centralized bottlenecks.

\textbf{Server side load-monitoring:} Another approach to load-monitoring is to 
distribute the load-monitoring among the caching shard servers. 
In~\cite{hong2013understanding}, each caching server tracks its own hot-spots.
When the hotness of a key surpasses a certain threshold, this key is replicated to 
$\gamma$ caching servers and the replication decision is broadcast to all the 
front-end servers. Any further accesses on this hot key shall be equally 
distributed among these $\gamma$ servers. This approach
aims to distribute the workload of the hot keys among multiple caching
servers to achieve better load balancing. Cheng et al.~\cite{cheng2015memory}
extend the work in~\cite{hong2013understanding} to allow moving coarse-grain key 
cachelets (shards) among threads and caching servers. Our approach reduces the need for
server side load-monitoring. Instead, load-monitoring happens at the 
edge. This allows individual front-end servers to independently identify their 
local trends.

\textbf{Client side load-monitoring:} Fan et al.~\cite{fan2011small} 
theoretically show through analysis and simulation that a small cache 
in the client side can provide load balancing to \textit{n} caching servers by 
caching only \textit{O(n log(n))} entries. Their result provides the theoretical
foundations for our work. Unlike in~\cite{fan2011small}, our approach
does not assume perfect caching nor \textit{a priori} knowledge of the workload access 
distribution. Gavrielatos et al.~\cite{gavrielatos2018scale}
propose \textit{symmetric caching} to track and cache the hot-most items 
at every front-end server. Symmetric caching assumes that all front-end
servers obtain the same access distribution and hence allocates the same cache
size to all front-end servers. However, different front-end servers might serve
different geographical regions and therefore observe different access distributions.
\algoname discovers the workload access distribution independently at each front-end server and adjusts the cache size to achieve a target load-imbalance $I_t$.
NetCache~\cite{jin2017netcache} uses programmable switches to implement 
heavy hitter tracking and caching at the network level. Like symmetric caching,
NetCache assumes a fixed cache size for different access distributions. To the best
of our knowledge, \algoname is the first front-end caching algorithm that exploits
the cloud elasticity allowing each front-end server to independently reduce the 
necessary required front-end cache memory to achieve back-end load-balance.

Other works in the literature focus on maximizing cache hit rates
for fixed memory sizes. Cidon et 
al.~\cite{cidon2015dynacache,cidon2016cliffhanger} 
redistribute available memory among memory slabs to maximize memory
utilization and reduce cache miss rates. Fan et al.~\cite{fan2013memc3}
use cuckoo hashing~\cite{pagh2004cuckoo} to increase memory utilization.
Lim et al.~\cite{lim2014mica} increase memory locality by assigning requests that 
access the same data item to the same CPU. Bechmann et al.~\cite{bechmann2018lhd}
propose Least Hit Density (LHD), a new cache replacement policy. LHD predicts
the expected hit density of each object and evicts the object with the lowest
hit density. LHD aims to evict objects that contribute low hit rates with respect to 
the cache space they occupy. 
Unlike these works, \algoname does not assume
a static cache size. In contrast, \algoname maximizes the hit rate of the 
available cache and exploits the cloud elasticity allowing front-end servers
to independently expand or shrink their cache memory sizes as needed.

\section{Conclusion} \label{sec:conclusion}

In this paper, we present \textit{Cache on Track }(\algoname), a decentralized, 
elastic, and predictive cache at the edge of a distributed cloud-based 
caching infrastructure. \algoname proposes a new cache
replacement policy specifically tailored for small front-end
caches that serve skewed workloads. Using \algoname, 
system administrators do not need to statically specify cache size 
at each front-end in-advance. Instead, they specify 
a target back-end load-imbalance $I_t$ and \algoname dynamically adjusts 
front-end cache sizes to achieve $I_t$. Our experiments show that \algoname's
replacement policy outperforms the hit-rates of LRU, LFU, ARC, and LRU-2 for the 
same cache size on different skewed workloads. \algoname achieves a target server size 
load-imbalance with 50\% to 93.75\% less front-end cache in comparison to other 
replacement policies. Finally, our experiments show that \algoname's resizing 
algorithm  successfully \textbf{auto-configures} 
front-end tracker and cache sizes to 
achieve the back-end target load-imbalance $I_t$ in the presence of workload distribution changes.

\balance
\bibliographystyle{abbrv}
\bibliography{main}

\end{document}